\documentclass[aps,graphicx,prl,twocolumn,groupedaddress]{revtex4}
\usepackage[dvips]{graphics}
\usepackage[dvips]{graphicx}
\begin{document}
\title{Theory of indirect resonant inelastic  X-ray scattering}
\author{Jeroen van den Brink$^1$ and Michel van Veenendaal$^{2}$}

\affiliation{
$^1$Institute-Lorentz for Theoretical Physics,  Universiteit  Leiden,
P.O. Box 9506, 2300 RA Leiden,
The Netherlands\\
$^2$Department of Physics, Northern Illinois University, De Kalb, Illinois 60115 and\\
Argonne National Laboratory, 9700 South Cass Avenue, Argonne, Illinois 60439
}
\date{\today}
\begin{abstract}
We express the cross section for indirect resonant inelastic X-ray scattering in terms of an intrinsic dynamic correlation function of the system that is studied with this technique. The cross section is a linear combination of the charge response function and the dynamic longitudinal spin density correlation function. This result is asymptotically exact for both strong and weak local core-hole potentials. We show that  one can change the relative charge and spin contribution to the inelastic spectral weight by varying the incident photon energy.
\end{abstract}
\maketitle
\narrowtext

{\it Introduction.} The probability for X-rays to be scattered from a solid state system can be enhanced by orders of magnitude when the energy of the incoming photons is in the vicinity of an electronic eigenmode of the system.  Such resonant X-ray scattering experiments~\cite{Kotani01} are performed on  e.g. the K-edges of transition metal ions, where the frequency of the X-rays is tuned to match the energy of an atomic $1s$-$4p$  transition, which is around 5-10 keV~\cite{Hasan00,Kim02,Hill98,Isaacs96,Kao96,Inami03,Abbamonte99}. At this resonant energy a $1s$ electron from the inner atomic core is excited into an empty $4p$ state, see Fig. 1. In transition metal systems the  empty $4p$ states are far (10-20 eV) above the Fermi-level, so that the X-rays do not cause direct transitions of the $1s$ electron into the lowest $3d$-like  conduction bands of the system. Still this technique is sensitive to excitations of electrons near the Fermi-level. The Coulomb potential of the $1s$ core-hole causes e.g. very low energy electron-hole excitations in the valence/conduction band: the core-hole potential is screened by the valence electrons.  When the excited $4p$-electron recombines with the $1s$ core-hole and the outgoing photon is emitted, the system can therefore be left behind in an excited final state.  Experimentally the momentum ${\bf q}$ and energy $\omega$ of the elementary excitation is determined  from the difference in energy and momentum between incoming and outgoing photons. Since the excitations are caused by the core-hole, we refer to this scattering mechanism as {\it indirect} resonant inelastic X-ray scattering (RIXS). 
 
RIXS is rapidly developing due to the recent increase in brilliance of the new generation synchrotron X-ray sources. High flux photon beams with energies that are tunable to resonant edges are now becoming available. At present an energy resolution of about 300 meV can be reached. In the near feature it seems experimentally feasible for RIXS to become sensitive to the low energy excitations of the solid, where excitation energies are of the order of room temperature.  Such low-lying electronic excitations can, for example, be collective features such as plasmons, magnons, orbitons, excitons and single-particle-like continua related to the band structure. RIXS provides a new tool to study these elementary excitations.

\begin{figure}
\includegraphics[width=0.65\columnwidth]{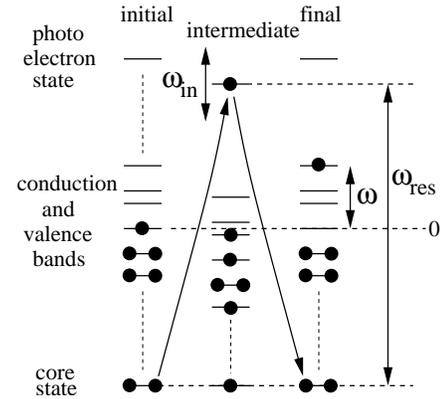} 
\caption{Schematic representation of the indirect resonant inelastic X-ray scattering (RIXS) process.}
\label{fig:scheme}
\end{figure}

For the interpretation of spectroscopic data, it is very important to express the scattering cross section for a technique in terms of physical correlation functions.  It is well known, for example, that in standard optical experiments the charge density response function (or dielectric function) at  ${\bf q} \simeq 0$ is measured. In electron energy loss experiments this linear response function is measured at finite transferred momentum. 

In this paper, we determine the dynamical correlation function that is measured in indirect resonant inelastic X-ray scattering. For a local core-hole potential, the dynamical correlation function turns out to be a linear combination of the charge density and longitudinal spin density response function. We show that the actual linear combination that is measured depends on the energy of the incoming photons and we determine the precise energy dependence of its coefficients. The expressions for the scattering cross section that we derive are asymptotically exact in the two limits of either a very strong or a very weak local core-hole potential. Our results, moreover, constitute a smooth interpolation between these two extremes.

{\it Series expansion of the scattering cross section.}
The Kramers-Heisenberg formula~\cite{Kramers25,Platzman69} for the resonant X-ray scattering cross section is~\cite{Note1}
\begin{eqnarray}
\left. \frac{ {\rm d}^2 \sigma} { {\rm d} \Omega {\rm d} \omega}  \right|_{res} &=& 
const \sum_{ f} | A_{fi} |^2 \ \delta (\omega-\omega_{fi}),
\label{eq:Kramers}
\end{eqnarray}
where $f$ and $i$ denote the final and initial state of the system, respectively. The sum is over all final states.
The momentum and energy of the incoming/outgoing photons is ${\bf q}_{{\rm in}/{\rm out}}$ and $\omega^0_{{\rm in}/{\rm out}}$ and
the loss energy $\omega= \omega^0_{\rm out}-\omega^0_{\rm in}$ is equal to the energy difference between the final 
and initial state $\omega_{fi}=E_f-E_i$. In the following we will take the energy of the initial state as reference energy: $E_i \equiv 0$. 
The scattering amplitude $A_{fi}$ is given by 
\begin{eqnarray}
A_{fi} = \omega_{\rm res} \sum_{ n} 
\frac{\langle f | \hat{O} | n \rangle \langle n | \hat{O} | i \rangle }
{\omega_{\rm in}-E_n -i \Gamma}, 
\label{eq:amplitude}
\end{eqnarray}
where $\omega_{\rm res}$ the resonant energy, $n$ denotes the intermediate states and  
$\hat{O}$ the (dimensionless) dipole operator that describes the excitation from initial
to intermediate state and the de-excitation from intermediate to final state. 
The energy of the incoming X-rays with respect to the resonant energy is $\omega_{\rm in}$ (this energy can thus
either be negative or positive: $\omega_{\rm in}= \omega^0_{\rm in}-\omega_{\rm res}$) and $E_n$ is the 
energy of intermediate state $|n\rangle$ with respect to the resonance energy. 

In the intermediate state a core-hole and a photo-excited electron are present.  
When we take the Coulomb interaction between the intermediate state core-hole and the valence band electrons 
into account, we obtain a finite inelastic scattering amplitude. In that case there is a non-zero probability that 
an electron-hole excitation is present in the final state, see Fig.1. 

The intermediate state, however, is not a steady state. The highly energetic $1s$ core-hole quickly decays e.g.
via Auger processes and the core-hole life-time is very short. The Heisenberg time-energy 
uncertainty relationship then implies that the core-hole energy has an appreciable uncertainty. This
uncertainty appears in the formalism above as the core-hole energy broadening $\Gamma$ which is
proportional to the inverse core-hole life-time. 
Note that the life-time broadening only appears in the intermediate states and not in the final or initial 
states as these both have very long life times. This implies that the core-hole broadening does not present an 
intrinsic limit to the experimental resolution of RIXS : the loss energy $\omega$ is completely determined by kinematics.

When the incoming energy of the X-rays is equal to a resonant energy of the system $\omega_{in} - E_n =0$ and we see 
from Eqs.~(\ref{eq:Kramers},\ref{eq:amplitude}) that the resonant enhancement 
of the X-ray scattering cross section is $(\omega_{\rm res}/\Gamma)^2$, which is $\sim 10^6$ for a transition 
metal K-edge~\cite{Blume85}. 

In a resonant scattering process, the measured system  is generally strongly perturbed. Formally this is clear from the Kramers-Heisenberg formula~(\ref{eq:Kramers}), in which  both the energy and the wavefunction of the intermediate state --where a potentially strongly perturbing core-hole is present-- appear. This is in contrast with canonical optical/electron energy loss experiments, where the probing photon/electron presents a weak perturbation to the system that is to be measured. 

To calculate RIXS amplitudes, one possibility is to numerically evaluate the Kramers-Heisenberg  expression. To do so, all initial, intermediate and final state energies and wavefunctions need to be known exactly, so that in practice a direct evaluation is only possible for systems that, for example, consist  of a small cluster of atoms~\cite{Tohyama02}. 
In this paper, however, we show that under the appropriate conditions we can integrate out the intermediate states from the Kramers-Heisenberg expression. After doing so, we can directly relate RIXS amplitudes to linear charge and spin response functions of the unperturbed system. For non-resonant scattering, one is familiar with the situation that the scattering intensity is proportional to a linear response function, but for a resonant scattering experiment this is a quite unexpected result. 

Let us proceed by formally expanding the scattering amplitude in a power series
\begin{eqnarray}
A_{fi} = \frac{ \omega_{\rm res} } {\omega_{\rm in} - i \Gamma}  
\sum_{l=0}^{\infty} M_l, 
\label{eq:A_fi}
\end{eqnarray}
where we introduced the matrix elements
\begin{eqnarray}
M_l= \sum_{ n} \left( \frac{E_n}{\omega_{\rm in}-i\Gamma} \right)^l
\langle f | \hat{O} | n \rangle \langle n | \hat{O} | i \rangle. 
\label{eq:M_l}
\end{eqnarray}
We denote the denominator of the expansion parameter $\omega_{\rm in}-i\Gamma$ by the complex number 
$\Delta$, so that
\begin{eqnarray}
M_l &=& \frac{1}{\Delta^l} 
\sum_{ n} \langle f | \hat{O}| n \rangle (E_n)^l \langle n | \hat{O} | i \rangle \nonumber \\
    &=& \frac{1}{\Delta^l} \langle f | \hat{O} (H_{\rm int})^l \hat{O} | i \rangle,
\label{eq:Mterms}
\end{eqnarray}
where $H_{\rm int}$ is the Hamiltonian in the intermediate state. We thus obtain the following series expansion for the resonant cross section
\begin{eqnarray}
\left. \frac{ {\rm d}^2 \sigma} { {\rm d} \Omega {\rm d} \omega}  \right|_{res} = 
const \sum_{ f} 
\left| \frac{\omega_{\rm res}}{\Delta} \sum_{l=0}^{\infty} M_l \right|^2  \delta (\omega-\omega_{fi}).
\label{eq:exp_cross}
\end{eqnarray}

{\it Spinless fermions.} We first calculate the resonant X-ray cross section in 
the case where the valence and conduction electrons are 
effectively described by a single band of spinless fermions: spin, and 
orbital degrees of freedom of the valence electron system are suppressed. Physically this
situation can be realized in a fully saturated ferromagnet.
 
The final and initial states of the system are determined by a Hamiltonian $H_0$ that
describes the electrons around the Fermi-level. The generic form of the full
many-body Hamiltonian is
\begin{eqnarray}
H_0 =   \sum_{i,j} t_{ij}  (c^{\tiny \dagger}_i c_j + c^{\tiny \dagger}_j c_i) +
c^{\tiny \dagger}_i c_i V_{ij} c^{\tiny \dagger}_j c_j,
\label{eq:H0}
\end{eqnarray}
where $i$ and $j$ denote lattice sites with lattice vectors ${\bf R}_i$ and ${\bf R}_j$ and
$i,j$ range from 1 to $N$, where $N$ is the number of sites in the system.
The hopping amplitudes of the valence electrons are denoted by $t_{ij}$ and the $c/c^{\tiny \dagger}$-operators
annihilate/create such electrons. 
The Coulomb interaction between valence electrons is  $V_{ij}= V_{|{\bf R}_i-{\bf R}_j|}$, as the
Coulomb interaction only depends on the relative distance between two particles.

The intermediate states are eigenstates of the Hamiltonian $H_{\rm int}= H_0 + H_{core}$, where $H_{core}$
accounts for the Coulomb coupling between the intermediate state core-hole and the valence electrons:  
\begin{eqnarray}
H_{core} = \sum_{i,j} s_i s^{\tiny \dagger}_i V^{c}_{ij} c^{\tiny \dagger}_j c_j,
\label{eq:Hcore}
\end{eqnarray}
where $s_i$ creates a core-hole on site $i$. 
We assume that the core-hole is fully localized
and has no dispersion. 
We will see shortly that this leads to major simplifications in the 
theoretical treatment of indirect RIXS. The core-hole -- valence electron interaction is attractive:  $V^{c} < 0$.
The dipole operators are given by
\begin{eqnarray}
\hat{O} = \sum_{i}  e^{-i {\bf q}_{\rm in}\cdot{\bf R}_i } s_i  p^{\tiny \dagger}_i + 
e^{i {\bf q}_{\rm out}\cdot{\bf R}_i } s^{\tiny \dagger}_i  p_i +h.c.,
\end{eqnarray} 
where $p^{\tiny \dagger}$ creates a photo-excited electron in a $4p$ state and $h.c.$ denotes the Hermitian conjugate of both terms.

In order to calculate the cross section, we need to evaluate the operator $(H_{\rm int})^l$ in equation (\ref{eq:Mterms}). A direct
evaluation of this operator is complicated by the fact that the initial and intermediate state Hamiltonians do not commute.
We therefore proceed by expanding  $(H_{\rm int})^l$ in a series
that contains the leading terms to the scattering cross section for both strong and weak core-hole potentials. After that we will 
do a full re-summation of that series.
We expand
\begin{eqnarray}
(H_{\rm int})^l \hat{O} | i \rangle
= (H_0+H_{core})^l   \hat{O} | i \rangle     \nonumber \\
\simeq \sum_{m=0}^{l} 
(H_{0})^m (H_{core})^{l-m} \hat{O}| i \rangle,
\label{eq:H_expand}
\end{eqnarray} 
where we used $[H_0,\hat{O}]=0$ and $H_0 | i \rangle \equiv 0$.
For strong core-hole potentials the expansion is dominated by the $m=0$ term $(H_{core})^l$.  Also for weak core-hole potentials 
we obtain the correct leading terms. It is straightforward to shown that for a strong core-hole potential the 
first correction to the expansion above is smaller by a factor $t/V^{c}$ and for a weak core-hole potential by a 
factor $V^{c}/t$. Thus the expansion is well behaved as in the appropriate limits the leading order corrections are vanishing. 

Let us for the moment consider the strong core-hole potential limit and keep in the expansion only the term $m=0$.  
Due to the fact that in the intermediate state only one core-hole is present and that this core-hole has no dispersion, 
we have
\begin{eqnarray}
M_l (V^{c}  \gg t)= \frac{1}{\Delta^l} \langle f | \sum_{i} e^{i {\bf q} \cdot{\bf R}_i }  
\left( \sum_{j} V^{c}_{ij} c^{\tiny \dagger}_j c_j \right)^l  | i \rangle, 
\label{eq:M_strong}
\end{eqnarray} 
where the transfered momentum ${\bf q} \equiv {\bf q}_{\rm out} - {\bf q}_{\rm in}$.

The first important observation is that the term $l=0$ does not contribute to the inelastic X-ray scattering intensity
because
$M_0=  \langle f | \sum_{i} e^{i {\bf q} \cdot{\bf R}_i }  | i \rangle = N \delta_{\bf q,0} \delta_{f,i}$, which only contributes to 
the elastic scattering intensity at ${\bf q}={\bf 0}$ and other multiples of the reciprocal lattice vectors. 
From inspection of equation (\ref{eq:M_l}) we see immediately that the $l=0$ term actually vanishes irrespective of the strength 
of the core-hole potential.
This is of relevance when we consider the scattering cross section in the so-called "fast-collision
approximation"~\cite{Veenendaal96}.  This approximation corresponds to the limit where the core-hole life 
time broadening is the largest energy scale in system ($\Gamma \rightarrow \infty$ or, equivalently,  
$Im [ \Delta ] \rightarrow -\infty$). In this limit only the $l=0$ term contributes to the indirect RIXS amplitude and 
 the resonant inelastic signal vanishes. In any theoretical treatment of indirect resonant scattering one therefore 
needs to go beyond the fast-collision approximation. 

The second observation is that $M_l$ is a $2^l$-particle correlation function. If we measure far away from resonance, 
where $|Re[\Delta]| \gg 0$, the scattering cross section is dominated by the $l=1$, two-particle, response function. When 
the incoming photon energy approaches the resonance, gradually the four, six, eight etc. particle 
response functions add more and more spectral weight to the inelastic scattering amplitude. Generally these 
multi-particle response functions interfere. We will show, however, that in the local core-hole approximation the multi-particle 
correlation functions in expansion (\ref{eq:H_expand}) collapse onto the dynamic two-particle (charge-charge) and 
four-particle (spin-spin) correlation function. 

In hard X-ray electron spectroscopies one often makes the approximation that the core-hole potential is local. This corresponds
to the widely used Anderson impurity approximation in the theoretical analysis of e.g. X-ray absorption and photo-emission, 
introduced in Refs.~\cite{Gunnarsson83,Kotani85,Zaanen86}.
This approximation is reasonable as the Coulomb potential is certainly largest on the atom where the core-hole is 
located. 

We insert a local core-hole potential $V^{c}_{ij}=U\delta_{ij}$ and find after performing the sum over $m$ in equation (\ref{eq:H_expand})
\begin{eqnarray}
M^{sf}_l =  \frac{1}{\Delta^l}  \frac{U^l-{E}_f^l}{1-E_f/U}
\langle f | \sum_{i} e^{i {\bf q} \cdot{\bf R}_i }  c^{\tiny \dagger}_i c_i  | i \rangle.
\label{eq:M_local}
\end{eqnarray} 
Using that 
$\sum_{i} e^{i {\bf q} \cdot{\bf R}_i }  c^{\tiny \dagger}_i c_i  =
\sum_{\bf k} c^{\tiny \dagger}_{{\bf k}+{\bf q}} c_{\bf k}  \equiv \rho_{\bf q}$  is the density
operator, we now perform the sum over $l$ in equation~(\ref{eq:A_fi}) and obtain
\begin{eqnarray}
A_{fi}^{sf}  =    P_1(\omega,U) \langle f| \rho_{\bf q} | i \rangle,  
\label{eq:A_spinless}
\end{eqnarray}
where  the resonant enhancement factor is
$P_1(\omega,U)  \equiv  U  \omega_{\rm res}  [(\Delta-U)(\Delta-\omega)] ^{-1}$ and $\omega=E_f$.
For spinless fermions with a local core potential the indirect RIXS cross section thus turns out
to be the density response function --a two-particle correlation function-- with a resonant prefactor $P_1(\omega)$ 
that depends on the loss energy $\omega$, the resonant energy ${\omega}_{\rm res}$,  on the distance from resonance $\omega_{\rm in}(=Re[ \Delta ]$), on the core-hole potential $U$ and on the core-hole life time broadening $\Gamma(=-Im [ \Delta ])$. 
The density response function is related to the dielectric function 
$\epsilon({\bf q},\omega)$ and the dynamic structure factor $S({\bf q},\omega)$~\cite{Mahan90}, so that for fixed  $U$ we obtain
\begin{eqnarray}
\left. \frac{ {\rm d}^2 \sigma} { {\rm d} \Omega {\rm d} \omega}  \right|_{res}^{sf} 
&\propto& -|P_1(\omega)|^2  \ {\rm Im} \ \left[ \frac{1}{V_{\bf q} \epsilon({\bf q},\omega)} \right] \nonumber \\
&\propto& |P_1(\omega)|^2  \ S({\bf q},\omega). 
\end{eqnarray} 

{\it Fermions with spin.} We generalize the calculation above to the situation where the electrons have an additional  
spin degree of freedom. In the Hamiltonians ~(\ref{eq:H0},\ref{eq:Hcore}) we now include a spin index 
$\sigma$ (with  $\sigma= \uparrow$ or $\downarrow$)
to the annihilation and creation operators: $c_i \rightarrow c_{i \sigma}$ and $c_j \rightarrow c_{j \sigma^{\prime}}$
and sum over these indices, taking into account that the hopping part of the Hamiltonian is diagonal in the spin variables.  
In order to re-sum the series in equation~ (\ref{eq:A_fi}) we now need to evaluate expansions of the number 
operators of the kind $(n_{\uparrow}+n_{\downarrow})^l$.
For $l >0$, this leads to
\begin{equation}
A_{fi}=
[P_1(\omega)  -P_2(\omega)]  \langle f|  {\bf S}_{\bf q}^2  | i \rangle   + P_2(\omega)  \langle f|\rho_{\bf q}  | i \rangle, 
\end{equation}
with $P_2(\omega,U) = P_1(2\omega,2U)/2 $ and  $\rho_{\bf q}  = \rho^{\uparrow} _{\bf q} + \rho^{\downarrow}_ {\bf q} $. Here
we also introduced the longitudinal spin density  correlation function  ${\bf S}_{\bf q}^2  \equiv \frac{1}{S(S+1)}\sum_{\bf k}  {\bf S_{k+q}} \cdot {\bf S_{-k} } $.
Clearly the contributions to the scattering rate from the dynamic longitudinal spin correlation function and the density correlation function need to be treated on equal footing as they interfere~\cite{Devereaux03-Kondo01}. 
Moreover, the spin and charge correlation functions have different resonant enhancements. For instance when $Re[\Delta]=U$, the scattering amplitude is dominated by $P_1(\omega)$ and hence by the longitudinal spin response function. At incident energies where $Re[\Delta]=2U$, on the other hand, $P_2(\omega)$ is resonating so that the contributions to the inelastic scattering amplitude of charge and spin are approximately equal.

Note that we only considered single band systems in this Letter. As a consequence the strength of the core-hole Coulomb potential is the same for all electrons in the valence band. In multi-band systems~\cite{Brink_tbp}, however, the Coulomb coupling will depend on the actual orbitals and bands that the electrons are in. For electrons in localized, atomic-like orbitals (e.g. transition metal $3d$ orbitals) the Coulomb interaction with the core-hole  is largest. Indirect RIXS is therefore mostly sensitive to excitations of these localized electrons. 

{\it Conclusions}.
We presented a series expansion of the indirect resonant inelastic X-ray scattering amplitude, which is asymptotically exact for both small and large local core-hole potentials. By re-summing the series, we find the dynamical charge and spin correlation functions that are measured in RIXS. The calculation of these correlation functions from e.g. model Hamiltonians or by means of {\it ab initio} electronic structure methods, will allow for a direct comparison of computed response functions with experimental RIXS spectra.

\begin{acknowledgments}
We thank J.P. Hill, S. Grenier and Y.J. Kim for stimulating discussions. 
We gratefully acknowledge support from the Argonne National Laboratory Theory Institute, Brookhaven National Laboratory (DE-AC02-98CH10996), the Dutch Science Foundation FOM (JvdB) and  the U.S. Department of Energy (DE-FG02-03ER460097) (MvV).
\end{acknowledgments}

\end{document}